\documentclass[10pt]{iopart}
\usepackage{bm}
\usepackage{color,graphicx}
\usepackage{tensor}
\begin{document}
\newcommand{\boldnabla}{\bm{\nabla}}


\title{Response to recent works on the discriminatory optical force for chiral molecules}

\author{Robert P Cameron}
\ead{r.cameron.2@research.gla.ac.uk}
\address{School of Physics and Astronomy, University of Glasgow,
Glasgow G12 8QQ, United Kingdom}

\author{Stephen M Barnett}
\address{School of Physics and Astronomy, University of Glasgow,
Glasgow G12 8QQ, United Kingdom}

\author{Alison M Yao}
\address{Department of Physics, University of Strathclyde,
Glasgow G4 0NG, United Kingdom}

\date{\today}


\begin{abstract}
\noindent We respond to recent works by Bradshaw and Andrews on the discriminatory optical force for chiral molecules, in particular to the erroneous claims made by them concerning our earlier work.
\end{abstract}


\maketitle



\section{Introduction}
\noindent We recently observed that the cycle-averaged, rotationally-averaged Lorentz force experienced by a small, non-magnetic, chiral molecule at rest or moving slowly in the presence of weak, far off-resonance, monochromatic light that is otherwise freely propagating can be cast as \cite{Cameron 14a, Cameron 14b}
\begin{eqnarray}
\langle\overline{\mathbf{F}}\rangle&=& a \boldnabla w \pm |b|\boldnabla h+ \dots
\end{eqnarray}
where $a=\alpha_{aa}(c|\mathbf{k}|)/3\epsilon_0$ is essentially a molecular polarizability which derives from the mutual interference of electric-dipole / electric-dipole transition moments within the molecule, $b=- \mu_0 c^2|\mathbf{k}| G_{aa}'(c|\mathbf{k}|)/3$ is an analogous quantity which derives instead from the mutual interference of electric-dipole / magnetic-dipole transition moments within the molecule, $w=\epsilon_0 \overline{\mathbf{E}\cdot\mathbf{E}}/2$ is the familiar electric energy density of the light and $h=\epsilon_0 c (\mathbf{A}^\bot\cdot\mathbf{B}-\mathbf{C}^\bot\cdot\mathbf{E})/2$ is the helicity density of the light \cite{Candlin,Ranada,Helicitypaper}. The plus and minus signs here distinguish between opposite molecular enantiomers. The first term $a\boldnabla w$ is the familiar dipole optical force \cite{Ashkin 70,Ashkin 01}, which acts to accelerate the molecule in a manner governed by electric energy gradients in the light. The second term $\pm |b|\boldnabla h$ is new and acts to accelerate opposite molecular enantiomers in opposite directions, in a manner governed by helicity gradients in the light: this is our discriminatory optical force for chiral molecules \cite{Cameron 14a, Cameron 14b}. Its form was also recognised, \textit{independently}, by a number of other authors in their considerations of an isotropic chiral dipole of unspecified constitution \cite{Canaguier 13, Wang 14, Ding 14}. We believe the contributions in \cite{Cameron 14a, Cameron 14b, Canaguier 13, Wang 14, Ding 14} to be mutually complimentary.
\newline\newline\noindent To date, we have identified a simple form of light sporting helicity fringes \cite{helicity fringes} for which $\boldnabla w=0$ but $\boldnabla h\ne0$ so that the force $\langle\overline{\mathbf{F}}\rangle=\pm |b|\boldnabla h+\dots$ is \textit{absolutely} discriminatory to leading order. Furthermore, we have suggested three devices based upon this, namely a chiral Stern-Gerlach deflector \cite{Cameron 14a, Cameron 14b}, a chiral diffraction grating \cite{Cameron 14a, Cameron 14b} and a discriminatory chiral diffraction grating \cite{Cameron 14b}, each with different functionalities. We have presented preliminary estimates to gauge the feasibility of these devices \cite{Cameron 14a, Cameron 14b} and are currently completing work on optimised, realistic designs and more precise calculations in which orientational effects and absorption are scrutinised in detail. This will be described by us elsewhere in due course.
\newline\newline\noindent We write in response to a number of recent publications authored by Bradshaw and Andrews \cite{1:1,question:question,2:3,3:2,5:4} on what is, essentially, our discriminatory optical force for chiral molecules. Our work \cite{Cameron 14a, Cameron 14b}, as well as the other founding contributions \cite{Canaguier 13, Wang 14, Ding 14} described above, is barely acknowledged by Bradshaw and Andrews except, that is, to make claims against it: in particular that helicity fringes \textit{do not} give rise to the effects predicted by us \cite{Cameron 14a,Cameron 14b}. In the present paper, we expose the flaws in the arguments presented by Bradshaw and Andrews and determine, as far as is possible, the origins of these. It is with regret that we do this, but it is necessary in order to prevent mistakes from propagating any further into the literature: the new field of discriminatory effects in the translational degrees of freedom of chiral objects \cite{Cameron 14a, Cameron 14b, Canaguier 13, Wang 14, Ding 14,Yong Li 07 b,Spivak 09,Li 10 a,Li 10 b,Jacob 12,Eilam 13,Donato 14,Tkachenko 14a,Tkachenko 14b,Canaguier 14,Chen 14,Alizadeh 15,Canaguier 15} holds much promise and it would be sad indeed to see it misdirected so early in its development. 


\section{The nature of an optical force}
\noindent Bradshaw and Andrews have expressed their view that dispersive optical forces derive from ``\textit{forward Rayleigh scattering}'' \cite{1:1,question:question,2:3,3:2,5:4}. Specifically, that these forces arise as the gradients of energy shifts, the forms of which follow from second-order perturbation theory:
\begin{equation}
\Delta W = \sum_{j\ne i } \frac{\langle i | \hat{H}_\textrm{int} | j\rangle \langle j | \hat{H}_\textrm{int} | i \rangle}{W_i-W_j}.
\end{equation}
Bradshaw and Andrews emphasise, in particular, their view that for ``\textit{the phenomenon of electromagnetic trapping, the initial and final states are identical ... In such circumstances, there is no net transfer of energy, or linear or angular momentum from the radiation field to the molecule}'' \cite{2:3} (see also \cite{1:1}). That this is \textit{manifestly} untrue is clear from a large body of experimental work that demonstrates the existence of an optical dipole force on isolated polarizable particles \cite{Ashkin 70, Ashkin 01}, including individual atoms in optical lattices \cite{Bloch 05}. It is incompatible, moreover, with the Lorentz force law.
\newline\newline\noindent In response to our particular observation that helicity fringes \cite{helicity fringes} yield an isolated and non-vanishing discriminatory optical force for chiral molecules \cite{Cameron 14a, Cameron 14b}, Bradshaw and Andrews write ``\textit{(a theoretical setup) proposed by Cameron et al. \cite{Cameron 14a,Cameron 14b} ... involves the scattering of overlapped beams of orthogonal linear polarization by chiral molecules. However, from a fundamental photonic perspective, such a system cannot relate to a chiral force (or a discriminatory energy) since the initial and final states are not identical \cite{Bradshaw 13}. This is readily demonstrated: since two beams are required for the proposed mechanism, and the process is elastic (no overall excitation of relaxation of molecular electronic state occurs) then any contributory mechanism has to involve a scattering event in which a photon is annihilated from one beam and a photon created into the other Since the initial and final radiation states of such a process are not identical, energy shifts cannot arise, and it becomes evident that there are no grounds for discriminatory forces to arise}'' \cite{question:question} (see also \cite{1:1}). That this line of reasoning is flawed is obvious from the fact that it does not give rise to the experimentally verified dipole optical force in an optical lattice, as discussed above. We shall show, moreover, that our discriminatory optical force for chiral molecules arises \textit{naturally} from the very formalism adopted by Bradshaw and Andrews.
\newline\newline\noindent The first conceptual error made by Bradshaw and Andrews derives from the fact that they have restricted their attention to a decomposition of the radiation field into plane wave modes (or similar travelling waves), together with energy eigenstates of the free radiation field in which each of these modes possesses a definite number of photons. The vital \textit{physical information} thereby neglected is that of the \textit{relative phase} between plane waves, which must be fixed rather than indeterminate in order to yield stable helicity fringes, an optical lattice etc. An example of an energy eigenstate of the free radiation field in which the relative phase of two plane waves \textit{is} fixed is \cite{Loudon,Cohen}
\begin{equation}
|i\rangle_\textrm{rad}=\sum_{n=0}^N \frac{ \sqrt{N!}}{ 2^{N/2}(N-n)! n! } (\hat{a}^\dagger_{\mathbf{k}_1\sigma_1})^{N-n} (\hat{a}^\dagger_{\mathbf{k}_2 \sigma_2})^n |0\rangle_\textrm{rad}
\end{equation}
where $|\mathbf{k}_1|=|\mathbf{k}_2|=|\mathbf{k}|$.  This describes the situation in which the standing wave mode 
\begin{eqnarray}
\nonumber
\frac{1}{\sqrt{2V}}\left[\tilde{\mathbf{e}}_{\mathbf{k}_1\sigma_1} \exp(\textrm{i}\mathbf{k}_1\cdot\mathbf{r})+\tilde{\mathbf{e}}_{\mathbf{k}_2\sigma_2} \exp(\textrm{i}\mathbf{k}_2\cdot\mathbf{r})\right]
\end{eqnarray}
possesses $N$ photons, satisfying
\begin{equation}
\hat{H}_\textrm{rad} |i\rangle_\textrm{rad}= (\hbar c |\mathbf{k}| N     +\mathcal{Z}_\textrm{vac}  ) |i\rangle_\textrm{rad}
\end{equation}
accordingly. Such modes are familiar, of course, from cavity quantum electrodynamics. Taking
\begin{eqnarray}
\mathbf{k}_1&=& |\mathbf{k}| (\sin\theta \hat{\mathbf{x}} +\cos\theta\hat{\mathbf{z}}), \nonumber \\
\mathbf{k}_2&=& |\mathbf{k}| (-\sin\theta \hat{\mathbf{x}} +\cos\theta\hat{\mathbf{z}}), \nonumber \\
\tilde{\mathbf{e}}_{\mathbf{k}_1\sigma_1}&=& \cos\theta \hat{\mathbf{x}}-\sin\theta \hat{\mathbf{z}}, \nonumber \\
\tilde{\mathbf{e}}_{\mathbf{k}_2\sigma_2}&=& \hat{\mathbf{y}},
\end{eqnarray}
we find using perturbation theory that the rotationally-averaged energy shift is
\begin{eqnarray}
\langle \Delta W\rangle&=&-\frac{\hbar c |\mathbf{k}| N \alpha_{aa}(c|\mathbf{k}|)}{6\epsilon_0 V} \nonumber  \nonumber \\
&&- \frac{\hbar  |\mathbf{k}| N G_{aa}'(c|\mathbf{k}|)}{3 \epsilon_0 V} \cos^2\theta \sin (2 |\mathbf{k}|\sin\theta X) \nonumber \\
&&+\dots
\end{eqnarray}
which is the result already reported by us on the basis of a semiclassical calculation\footnote{We did in fact note in \cite{Cameron 14a} that ``the (form) of ($\langle\Delta W\rangle$) can ... be justified by an appropriate calculation in the quantum domain".} \cite{Cameron 14a}. The gradient with respect to $X$ of this $\langle \Delta W\rangle$ yields the isolated and non-vanishing discriminatory optical force for chiral molecules already described in \cite{Cameron 14a, Cameron 14b}. Thus, we have shown that the methods favoured by Bradshaw and Andrews, when applied correctly, lead \textit{directly} to the disputed helicity gradient force \cite{Cameron 14a, Cameron 14b}.


\section{Uniqueness of predictions}
\noindent There are, unfortunately, further problems with the work of Bradshaw and Andrews and it be would remiss of us not to address these.
\newline\newline\noindent Bradshaw and Andrews begin their investigations by considering an orientated molecule together with a single circularly polarised plane wave mode (or similiar). The corresponding energy shift reported by them \cite{1:1, 2:3}, when written in the standard notation \cite{Barron}, is
\begin{eqnarray}
\Delta W&=& -\frac{\hbar c |\mathbf{k}| n_{\mathbf{k}\sigma}}{2\epsilon_0 V} \tilde{e}_{\mathbf{k}\sigma a}^\ast \tilde{e}_{\mathbf{k}\sigma b}\Big( \tilde{\alpha}_{ab}(c|\mathbf{k}|) \nonumber \\
&&  + \frac{k_c}{c|\mathbf{k}|} [\epsilon_{dca}\tilde{G}_{bd}^\ast(c|\mathbf{k}|)+\epsilon_{dcb}\tilde{G}_{ad}(c|\mathbf{k}|)] \nonumber \\
&&+\frac{\textrm{i} k_c}{3c}\big\{ \frac{1}{2}[\tilde{A}_{abc}(c|\mathbf{k}|) +\tilde{A}_{abc}^\ast (c|\mathbf{k}|) ]  \nonumber \\
&&  -\frac{1}{2}[\tilde{A}_{bac}(c|\mathbf{k}|) +\tilde{A}_{bac}^\ast (c|\mathbf{k}|)] \big\}\Big). 
\end{eqnarray}
Bradshaw and Andrews then focus their attention upon real molecular wavefunctions, as is usual, but \textit{neglect} electric quadrupole contributions on the grounds that $\tilde{A}_{abc}(c|\mathbf{k}|)$ is real. The form for $\Delta W$ thus obtained by them constitutes the basis for their subsequent investigations, in which they give particular attention to oriented effects.
\newline\newline\noindent Unfortunately, the form reported by Bradshaw and Andrews for $\Delta W$, as seen above, is \textit{incorrect} and their neglect of electric-quadrupole contributions cannot be justified. Consequently, the results seen in equations (10), (13), (15), (16), (17), (18) and (19) of \cite{1:1}; (4), (5), (7) and (8) of \cite{question:question}; (10), (11), (12), (13), (17), (19) and (21) of \cite{2:3} and (1) and (2) of \cite{5:4} are in error, as they are not uniquely defined: their forms change with changes in the location chosen for the origin of the molecular multipole expansion. An energy shift and associated force \textit{cannot} take values that depend on the manner in which we choose to calculate them. Analogous omissions are seen also in their recent works on optical binding \cite{5:4,4:5}.
\newline\newline\noindent The correct result has already been reported elsewhere by one of us \cite{Rob PhD} and is
\begin{eqnarray}
\Delta W&=& -\frac{\hbar c |\mathbf{k}| n_{\mathbf{k}\sigma}}{2\epsilon_0 V} \tilde{e}_{\mathbf{k}\sigma a}^\ast \tilde{e}_{\mathbf{k}\sigma b}\Big( \tilde{\alpha}_{ab}(c|\mathbf{k}|) \nonumber \\
&&+ \frac{k_c}{c|\mathbf{k}|} [\epsilon_{dca}\tilde{G}_{bd}^\ast(c|\mathbf{k}|)+\epsilon_{dcb}\tilde{G}_{ad}(c|\mathbf{k}|)] \nonumber \\
&&+\frac{\textrm{i} k_c}{3c}\big\{ \tilde{A}_{abc}(c|\mathbf{k}|)-\tilde{A}_{bac}^\ast (c|\mathbf{k}|) \big\}\Big)+\dots \nonumber \\
&=&-\frac{\hbar c |\mathbf{k}| n_{\mathbf{k}\sigma}}{2 \epsilon_0 V} \tilde{e}_{\mathbf{k}\sigma a}^\ast \tilde{e}_{\mathbf{k}\sigma b} \left[\tilde{\alpha}_{ab} (c|\mathbf{k}|) +\tilde{\zeta}_{abc}(c|\mathbf{k}|) k_c/|\mathbf{k}|\right] + \dots .
\end{eqnarray}
This form can be checked by noting, for example, that it is manifestly independent of the location of the origin of the molecular multipole expansion, as it must be \cite{Barron}. We observe, moreover, that this form leads to the correct expressions for the refractive indices and optical rotation angles of various media \cite{Rob PhD}.
\newline\newline\noindent As is well-established in studies that involve orientated molecules, $\tilde{G}_{ab}(c|\mathbf{k}|)$ and $\tilde{A}_{abc}(c|\mathbf{k}|)$ must be considered \textit{together}, in general, in order to obtain unique and hence physically meaningful predictions \cite{Barron}. As Barron wrote, ``\textit{It is emphasised that the separate electric dipole-magnetic dipole and electric dipole-electric quadrupole contributions are origin dependent in an oriented sample: the change in one contribution on moving the origin is cancelled by the change in the other. Consequently, the analysis of optical rotation or circular dichroism data on oriented systems can be quite wrong if only the electric dipole-magnetic dipole contribution is considered.}'' \cite{Barron}. Bradshaw and Andrews have made precisely this mistake.


\section{Numerical estimates}
\noindent We conclude by commenting on the viability of the scheme put forward by Bradshaw and Andrews for the separation of opposite molecular enantiomers. They have suggested that illuminating an ordinary fluidic sample of chiral molecules, perhaps in the presence of a static electric field, with circularly polarised light will encourage an appreciable difference in concentrations of opposite molecular enantiomers \cite{1:1,question:question,2:3,3:2,5:4}. Various numerical estimates are reported by Bradshaw and Andrews \cite{1:1,question:question,2:3,3:2,5:4} in this regard. Here we simply note that Bradshaw and Andrews ``\textit{define}'' 
\begin{equation}
G'_{aa}(c|\mathbf{k}|)/c=4 \pi \epsilon_0 d^3/ 137,
\end{equation}
where $d$ is a characteristic molecular dimension \cite{question:question}. Using the enormous value\footnote{A cube of edge $d=10^{-8}$m could encompass some $10^6$ atoms of notional width $10^{-10}$m, say.}  of $d=10^{-8}$m proposed by them \cite{1:1,question:question}, we find that the expression quoted above yields $G_{aa}'(c|\mathbf{k}|) /c=10^{-36}$kg$^{-1}$.s$^4$.A$^2$, with an opposite sign, we must presume, for the opposite molecular enantiomer. We note, for comparison, that in our estimates for hexahelicene, a real molecule with a very large chiroptical response (which appears, in fact, in Bradshaw and Andrews's figures \cite{2:3,3:2}), $G_{aa}'(c|\mathbf{k}|)/c=\pm 3\times 10^{-43}$kg$^{-1}$.s$^4$.A$^2$ at an angular frequency of $c|\mathbf{k}|=2\times 10^{15}$ s$^{-1}$, where the plus and minus signs refer, respectively, to the left- and right-handed helical forms of the molecule \cite{Cameron 14a, Cameron 14b}. The reader will note a discrepancy of \textit{seven orders of magnitude}. This stems in part from their choice of $d=10^{-8}$m, which in reality would correspond to a molecule that possesses thousands, tens of thousands or perhaps even millions of degrees of conformational freedom. The very notion of having clearly defined opposite molecular enantiomers, or far off-resonant light for that matter, is difficult to believe for such enormous molecules.


\section{Conclusion}
\noindent We have shown that the molecular QED methods favoured by Bradshaw and Andrews lead directly to the chiral force predicted by us \cite{Cameron 14a, Cameron 14b}.  Far from refuting our proposal, their methods (when correctly applied) reproduce our own predictions.
We have determined the source of error in their work and, moreover, corrected two further slips: the omission of electric quadrupole interactions and the proposal of an optical activity polarizability that exceeds the largest accessible values by many 
orders of magnitude. 
\newline\newline\noindent In view of the above, and more besides \cite{1:1, question:question, 2:3,3:2,5:4}, we put it to the reader that it is, in fact, the claims made by Bradshaw and Andrews that are ``\textit{extravagant}'' \cite{question:question}.


\section{Acknowledgements} This work was supported by the Engineering and Physical Sciences Research Council grants EP/101245/1 and EP/M004694/1.



\end{document}